\documentclass[conference]{IEEEtran}
\IEEEoverridecommandlockouts
\usepackage{cite}
\usepackage{url}
\usepackage{amsmath,amssymb,amsfonts}
\usepackage{algorithmic}
\usepackage{graphicx}
\usepackage{textcomp}
\usepackage{xcolor}
\def\BibTeX{{\rm B\kern-.05em{\sc i\kern-.025em b}\kern-.08em
    T\kern-.1667em\lower.7ex\hbox{E}\kern-.125emX}}
\begin{document}

\title{GraphQL Live Querying with DynamoDB\\
}

\author{\IEEEauthorblockN{Austin Silveria}
\IEEEauthorblockA{
\textit{California Polytechnic State University, San Luis Obispo}\\
San Luis Obispo, United States \\
silveria@calpoly.edu}
}

\maketitle

\begin{abstract}
We present a method of implementing GraphQL live queries at the database level. Our DynamoDB simulation in Go mimics a distributed key-value store and implements live queries to expose possible pitfalls. Two key components for implementing live queries are storing fields selected in a live query and determining which object fields have been updated in each database write. A \verb|stream(key, fields)| request to the system contains fields to include in the live query stream and on subsequent \verb|put(key, object)| operations, the database asynchronously determines which fields were updated and pushes a new query “view” to the stream if those fields overlap with the \verb|stream()| request. Following a discussion of our implementation, we explore motivations for using live queries such as simplifying software communication, minimizing data transfer, and enabling real-time data and describe an architecture for building software with GraphQL and live queries.
\end{abstract}


\section{Introduction}
Over the past decade, the scale of internet applications, and by effect their infrastructure, have grown exponentially. Massively distributed NoSQL databases such as DynamoDB \cite{b1}, Cassandra \cite{b2}, and BigTable \cite{b3} have risen in popularity due to their horizontal scalability and fault tolerance compared to traditional relational databases. As discussed in the DynamoDB paper \cite{b1}, primary design goals of web scale data storage solutions are incremental scalability (scale out one node at a time), symmetry (each node should perform the same function), decentralization (favor peer-to-peer techniques over centralized control), and heterogeneity (nodes can have different capacities). A comparison of DynamoDB, Cassandra, and BigTable addresses this in more depth \cite{b4}--DynamoDB and Cassandra adopt all of these principles in their design, but BigTable opts for a centralized control mechanism instead of being completely symmetric and peer-to-peer. In the context of the CAP theorem, DynamoDB and Cassandra sacrifice strong consistency in favor of high availability and partition tolerance, while BigTable sacrifices some availability for strong consistency and partition tolerance. Cassandra and BigTable share the same data model--a multi-dimensional map--and local persistence strategy--writing to an in-memory commit log, persisting it to disk on overflow, and asynchronously compacting the persisted commit logs. On the other hand, DynamoDB favors a durable and simple key-value store optimized for reads at the cost of lower write throughput due to the necessity of writing to disk rather than an in-memory buffer. To scale horizontally, BigTable dynamically partitions row ranges of the table based on sort order while DynamoDB and Cassandra use consistent hashing \cite{b5}.

As the quantity of data has grown, the complexity of application data models has grown as well. GraphQL \cite{b6} is a query language introduced by Facebook to simplify large data models by unifying multiple APIs under a single “object” interface and allowing clients to specify slices of data they want. Figure 1 illustrates the flexibility of GraphQL--clients can select subsets of models defined within the schema and GraphQL will only return that data from the server. Compared to REST, this model of interacting with a single, unified graph has greatly improved developer experience in working with large data models.

\begin{figure*}[htbp]
\centerline{\includegraphics{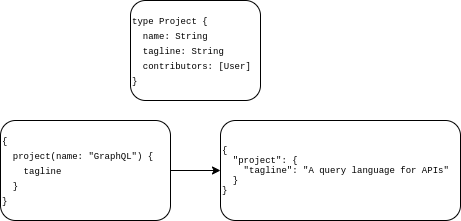}}
\caption{Example of simple GraphQL query selecting a subset of object fields.}
\label{fig:gql}
\end{figure*}

Beyond dealing with vast quantities of data and complex data models, internet applications also face the problem of asynchronously fetching and transforming data. When maintaining up to date state within application components, developers must be wary of race conditions, constantly re fetch new data, and transform data, all from different models. Streaming mechanisms such as DynamoDB streams \cite{b7} and GraphQL subscriptions have proved powerful methods of maintaining fresh data with ease--each time there is an update, receive the new data. However, DynamoDB streams are limited in their functionality: it is only possible to stream all updates from a table and in one of four formats--keys only (the key attributes of the modified item), new image (the entire item after it is modified), old image (the entire item before it is modified), or new and old images (both new and old items). And GraphQL subscriptions must be tied to specific events such as another GraphQL mutation--they are not live queries in the sense that the subscription is not fired every time the query would result in something different.

In this paper, we present a method of supporting streamed live queries at the database level. Specifically, we implement live queries in a simplified, simulated version of DynamoDB using the Go programming language\footnote[1]{Our code is available at: \url{https://github.com/austinsilveria/LiveGraphQL}}, but our method can also be implemented in other data storage solutions such as Cassandra or BigTable. We adopt GraphQL’s idea of selecting a subset of an object and push a new update to a stream any time a write to the key-object store contains a new value for one of the selected fields.

In the following section we discuss the implementation of our DynamoDB simulation and corresponding live query implementation. We then discuss the results of implementing live queries at the database level and what could be built on top of this functionality. Finally, we identify areas of future work and present our conclusions.

\section{Implementation}

In supporting live queries, it is necessary to store the selection of fields to listen to updates on and be able to determine which fields are updated on each write. Updating a requested live query can be done asynchronously on writes to avoid affecting the performance of current database operations.

Our implementation simulates DynamoDB as presented in the 2006 paper \cite{b1} by leveraging Goroutines. At a high level, separate Goroutines are spawned and act as physical nodes in the network--each Goroutine manages multiple virtual nodes, exchanges membership information with a gossip based protocol, and stores data in in-memory maps based on consistent hashing \cite{b5}. The simulation focuses on supporting the core aspects of DynamoDB necessary to introduce live querying and therefore we leave version reconciliation and optimized replica synchronization to a production implementation.

The interface is the same \verb|get(key)| and \verb|put(key, object)|, but with an additional operation for streaming a live query: \verb|stream(key, fields)|. All operations use an MD5 hash function to identify the nodes responsible for a specific key based on continuous hashing. The \verb|stream(key, fields)| operation sends a request to the nodes responsible for a key to store the given stream request with the object. On subsequent \verb|put()| operations, if a new value is present for any of the fields contained in the stored stream request, a new “view” of the fields will be pushed to the stream. Our implementation does not hook up the stream to a client, but in practice any message queue such as SNS/SQS \cite{b8} or NSQ \cite{b9} can be used.

To determine if the fields requested by a particular \verb|stream()| are updated in a \verb|put()|, our simulation uses an \verb|update(key, sparseObject)| interface for our put execution. The client only includes the fields they wish to update in the request and the object is merged with the currently stored version before being persisted. This allows an easy method of determining whether to push the selection of the updated object to the stream because the fields that are updated are explicitly defined in the request--they can be compared against the fields selected by the stream request that is stored with the object. However, this degrades write performance because an object must be read and merged with the update before the write can complete.

\begin{figure*}[htbp]
\centerline{\includegraphics{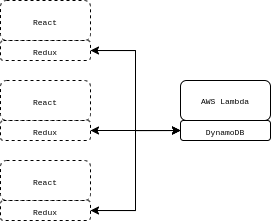}}
\caption{Two way live query between states of frontend and backend after frontend client has initiated a connection with the backend service.}
\label{fig:twoway}
\end{figure*}

Two options arise to solve the problem of efficiently determining which fields have been updated. The first is persisting the sparse updates immediately, completing the write, and asynchronously merging the versions later. This would improve the write performance and still allow easy access to updated fields, but could cause higher read latencies if requests come in before the versions are merged. Also, fields could be included in the update, but not actually contain different values. The second option uses the full object \verb|put(key, object)| interface, stores the new version of the object alongside the old version (as is already being done by DynamoDB based on the paper), completes the write, and asynchronously diffs the two versions to determine which fields were updated. This option keeps both read and write latencies low, but increases computation cost from computing the diff and increases the stream latency from when the update occurred. To reduce the amount of computation needed to detect which fields have different values between two versions, it is possible to use Merkle trees \cite{b10} in the same way DynamoDB already does to minimize data transfer between replicas. The leaves of the hash tree are the hashes of each field value separately, and the parents are the hashes of two children together. If the roots of the two hash trees (one for each version) are different, the tree can be efficiently traversed to find the fields that are different. In practice, this method of using new and old versions is similar to DynamoDB’s existing streaming solution, but operates at the key level rather than the table level--the new and old versions are only subject to a diff if the object key is included in the stream request. AWS AppSync’s Delta Sync \cite{b11} operation is also based on DynamoDB’s versioning capabilities, but relies on the client to request updates rather than server streaming updates. Options for determining the updated fields on a write can be chosen based on the needs of each system.

Similar to DynamoDB, to implement live queries in Cassandra or BigTable it is necessary to store requested fields to stream and be able to determine which fields (in this case column families) are updated without degrading the performance of other operations. Multiple stream requests existing at once can be stored in a super-column and since Cassandra and BigTable have more structured access patterns than DynamoDB, it is possible to determine which columns are updated based on the specified row mutation. Beyond this simple intuition, we leave further exploration of extending live queries to additional data storage solutions to future work.

\begin{figure*}[htbp]
\centerline{\includegraphics{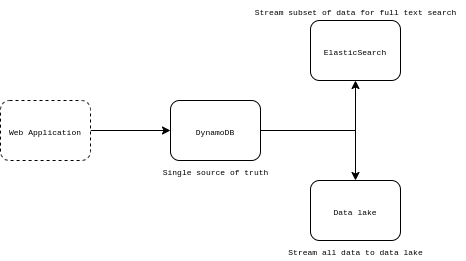}}
\caption{Derive new data stores by live querying a range of data in existing store.}
\label{fig:derive}
\end{figure*}

\section{Discussion}
Live queries allow decoupled software applications to efficiently and easily communicate subsets of state. Independent pieces of software such as backend services and frontend applications manage a subset of the organization’s “full scope” of state in different ways, and commonly rely on \verb|put()| and \verb|get()| operations to share data. This communication model is complicated by the need to poll for new data and the existence of multiple data models. GraphQL reduced the complexity of data models by unifying them under a single interface, minimized data transfer by allowing a subset of data to be retrieved, and reduced the need for polling with subscriptions. Subscriptions are fragile, however, because they are fired based on events rather than any time a query would change--in large applications it becomes intractable to specify all events that could cause a query to change. With live queries at the database level, the software is agnostic to events causing the query to change and can easily trigger updates when a query result changes.

Event-agnostic live queries enable subsets of state that are shared across multiple applications to remain in sync, simplifying state communication logic, minimizing data transfer, and providing real-time capabilities. For front end applications, live queries can keep a client’s cache hydrated with one request, only send delta updates as seen in AppSync’s Delta Sync \cite{b11}, and extend real-time capabilities to any piece of state. For backend services such as a data lake, machine learning system, or a dependent microservice, live queries can efficiently propagate new information to where it is needed with minimal operational overhead.

Consider the simplified framework of viewing software applications as two layers: logic and state. Frontend applications use logic to present a user interface based on their state and to apply updates to their state. Backend services use logic to present an API and to apply updates to their state. In the case of a social media post, its state may exist in multiple frontend instances of the web application (multiple users viewing the post) and one backend service as the source of truth--the states of the applications are “overlapping” on that post. With live queries, when any one of those applications applies an update to the “overlapping” state, it is possible to propagate the update to all other applications that “overlap” on that state. In other words, a frontend client can initiate a two-way live query with a backend service such that updates to “overlapping” state by the frontend client are live queried by the backend service, resulting in database writes; the backend service can then propagate these updates to other clients that have initiated a live query. This is essentially a declarative software communication model compared to the imperative \verb|put()| and \verb|get()| model. Figure 2 illustrates this concept using popular web technologies.

In the context of a network of backend services spanning multiple teams, GraphQL with live queries presents several opportunities to simplify software development. Support for new operations on data, such as a new type of query, may require a new data store to efficiently satisfy the requirements of the operation. This often results in the team owning the data having to implement the new data store and operation because they are domain experts in their “scope” of the organization’s data. With GraphQL, however, its strongly typed schema and introspection system can allow the client team to easily discover and learn about the owning team’s data. Coupled with live queries, the client team can effectively derive the new data store to support their operation by streaming the necessary scope of data from the original store. The derived data store uses a one-way live query to receive the data, apply transformations, and maintain it in a different format. With this case in mind, Figure 3 shows the possibilities of deriving multiple data stores serving different use cases from a single source of truth.

\section{Conclusion and Future Work}

We have described a method of implementing GraphQL live queries at the database level. The two necessary components for supporting live queries are storing the requested fields of the query and determining which fields of an object are updated on each write operation. A DynamoDB simulation written in Go implements basic live query functionality and illustrates the unique issues introduced by the problem. With the learnings from this exercise, we identify multiple options for supporting live queries in a distributed key-value store like DynamoDB without sacrificing performance or reliability of existing operations. We have also discussed motivating use cases for GraphQL live queries such as real-time web applications and simple backend service communication. A clear next step for future work is implementing live query functionality in a production-ready distributed database. With this in place, it will be possible to use a simpler software communication model and improve processes of data intensive software development.

\end{document}